\documentstyle[12pt,amssymb,psfig]{article}
\newcommand{\be}{\begin{equation}}
\newcommand{\ee}{\end{equation}}
\newcommand{\ea}{\end{array}}

\def\Frac#1#2{\frac{\displaystyle{#1}}{\displaystyle{#2}}}

\begin{document}

\begin{flushright}

FTUV/97-40

IFIC/97-40

UG-DFM-5/97
\end{flushright}

\begin{center}
{\Large {\bf Inclusive $(e,e^\prime N)$, $(e,e^\prime NN)$,
$(e,e^\prime \pi)$ ...
 reactions in nuclei}}
\end{center}

\vspace{0.5cm}

\begin{center}
{\large {A. Gil$^1$, J. Nieves$^2$ and E. Oset$^1$  }}
\end{center}

\vspace{0.3cm}

{\small {\it

$^1$Departamento de F\'{\i}sica Te\'orica and IFIC, Centro Mixto Universidad
de Valencia - CSIC, 46100 Burjassot (Valencia) Spain.

$^2$ Departamento de F\'{\i}sica Moderna, Universidad de Granada, 
18071 Granada, Spain.}} 

\vspace{0.7cm}

\begin{abstract}

We study the inclusive  $(e,e^\prime N)$, $(e,e^\prime NN)$,
$(e,e^\prime \pi)$, $(e,e^\prime \pi N)$ reactions in nuclei using
a Monte Carlo simulation method to treat the multichannel problem of the
final state. The input consists of reaction probabilities 
for the different steps 
evaluated using microscopical many body methods.

We obtain a good agreement with experiment in some channels where there is
data and make predictions for other channels which are presently 
under investigation in several electron
laboratories. The comparison of the theoretical
results with experiment for several kinematical conditions and diverse
channels can serve to learn about different
physical processes ocurring in the reaction. The potential of this 
theoretical tool to make prospections for possible experiments, aiming
at pinning down certain reaction probabilities, is also emphasized.

\end{abstract}

\vspace*{0.5cm}
\vspace*{0.5cm}

\section{Introduction}

Experiments on exclusive $(e,e^\prime )$ reactions in nuclei are becoming 
more and more common in present electron laboratories. Experiments 
concentrate on  $(e,e^\prime p )$,  $(e,e^\prime NN )$,  $(e,e^\prime \pi )$,
 $(e,e^\prime \pi N )$ etc. reactions which, together with the inclusive
  $(e,e^\prime )$ measurements should help us understand better the mechanisms
  of nuclear excitation by electromagnetic probes. Proton knockout reactions
  at low energies have been used as tools to learn about nuclear
  structure, high momentum components in the nucleus,
   nuclear correlations, etc. \cite{BOFFI,IRELAND,VIRLE,KESTER,POLLS}.
   Two nucleon emission reactions have been encouraged with the hope 
   that they can teach us 
   something about nuclear correlations \cite{VIRLE,GIUSTI,JAN}. However, meson
   exchange currents mechanisms and delta excitation are usually competitive
   and, together with complications in dealing with final state interactions, 
   make the extraction of information on nuclear correlations a difficult 
   task \cite{VIRLE}. 
   
               Mechanisms for 2N emission in $(e,e^\prime )$ reactions have been developed
  in \cite{PACATI,RADICI} and \cite{JAN2,JAN3}. The analogous reaction of two nucleon emission induced by real photons
  has received parallel attention \cite{GIANNINI, BOFFI2, RAFA}. The approaches
  of the Pavia \cite{PACATI, RADICI, BOFFI2} and Gent groups \cite{JAN2,JAN3}
  are suited to deal with low energy photons, below the pion emission threshold
  which opens new channels for 2N emission \cite{RAFA}. Both groups take
  into account final state interactions of the two nucleons with
  the nucleus: the Pavia group by means of nucleon distorted waves with a complex
  optical potential, while the Gent group uses a real 
  nucleon-nucleus potential. The distortion of the nucleon wave function by a complex
  optical potential removes all events where the nucleons 
  collide with other nucleons.
  The procedure is suited to study NN emission leading to specific final 
  nuclear states. However, if one is interested in ``inclusive'' 2N emission,
  meaning that one wishes to compare with an experiment where 2N are detected
  but one is implicitly summing over all final nuclear states, including
  more than 2N emission, one has to find a different scheme. The distortion
  by a real potential does not eliminate the events where there are nucleon
  collisions. On the other hand, it does not account either for events
  coming from one nucleon emission followed by secondary collisions of the
  nucleon which lead to two or more nucleon emission. Such events have been
  taken into account in ref. \cite{TAKAKI} although they are insufficient
  to account for the 2N spectra. This sort of final state interaction
  is also adressed in \cite{GIUSTI2} in order to account for contributions
  to   $(e,e^\prime pp )$ from  $(e,e^\prime pn )$  followed
  by charge exchange in a secondary collision of the neutron. However, only
  analogue intermediate states are allowed to be excited in that approach
  and in nuclei with $T=0$ there is no contribution from this channel.
       None of the above mentioned approaches deals explicitly with
       pion production and hence they miss the most important channel
       of 2N emission at energies above $E_{\gamma}=250\,MeV$ which consists
       of real pion creation followed by the absorption of the pion
        on its way out  the nucleus (indirect photon absorption) \cite{RAFA}.
        In ref. \cite{RAFA} a thorough work was done for $\gamma$ absorption
        in nuclei, including all possible many step processes:
        a) photon absorption by a pair of nucleons, b) photon absorption 
        by a trio of nucleons, c) $(\gamma,\pi)$ at a point in the nucleus followed
        by pion and nucleon collisions, d) $(\gamma,\pi)$ followed by pion
         absorption
        on its way out of the nucleus, etc. The work uses
       probabilities of photon absorption or pion photoproduction as input, 
        calculated with a microscopic many 
        body approach \cite{CARRASCO}. Then in a second step,
         it uses a Monte Carlo simulation approach to account  for the possible 
         reactions
which can take place in the interaction of the particles 
produced with the nucleus. The approach is rather succesful in reproducing the 
$(\gamma, pn)$ cross sections \cite{GROSS,GRAB,HELH} although it has
some difficulties in the $(\gamma, pp)$ channel which experimentally
shows a very small cross section. Similarly the agreement with the 
$(\gamma,\pi)$ \cite{ARENDS} and $(\gamma,\pi N)$ \cite{GROSS} channels
is also quite good \cite{LOREN}. 

The work of ref. \cite{CARRASCO} has been recently extended to virtual photons
in order to study the $(e,e^{\prime})$ inclusive reaction \cite{AMPARO}. 
On top of several subtleties of virtual photons,
the extension of the microscopic many body work of 
\cite{CARRASCO}  requires the study of the quasielastic channel,
$\gamma^* N \rightarrow N$, which is absent in real photons
(at photon energies bigger than 100 MeV, in practical terms). The work
of \cite{AMPARO} evaluates the inclusive $(e,e^{\prime})$ cross section
and at the same time all the probabilities for the reactions that take place
in the interaction of $\gamma^*$ with the nucleons in the first step:
$(\gamma^*, \pi)$, $(\gamma^*, N)$, $(\gamma^*,NN)$ and $(\gamma^*, NNN)$.

The agreement of the results of ref. \cite{AMPARO} with the inclusive
$(e,e^{\prime})$ data is quite good, but since the theory is able
to relate the strength of the cross section to different physical channels,
the theory has a higher potential predictive power than just the prediction
of the strength of the total cross section and it is the purpose of the present
paper to exploit this potential.

In the next section we summarize the technical details which are used in
\cite{AMPARO} in order to obtain the probabilities.

The present work hence aims to get a complete description of all processes
which can occur in the $(e,e^{\prime})$ reaction in nuclei in a range of energies 
between 100 MeV and about 600 MeV of the virtual photon. These are
typical energies of the Mainz facility.
In this way we shall evaluate cross sections for $(e,e^{\prime}N)$,
$(e,e^{\prime}NN)$, $(e,e^{\prime}\pi)$, $(e,e^{\prime} \pi N)$ etc.
reactions and will compare them with some experiments in order to interprete the data
and relate them to the relevant physical mechanisms of the reaction.

\section{The evaluation of probabilities in the first step}

In a previous paper \cite{AMPARO} we carried out a thorough evaluation
of the different mechanisms entering the $(e,e^\prime )$ inclusive
reaction. This allowed us to compare with the inclusive
cross section $d^2\sigma /d\Omega^{\prime} dE^{\prime}$ 
($\Omega^{\prime}$, $E^{\prime}$ solid angle and energy of the outgoing 
electron) and 
also with the longitudinal
and transverse response functions. The agreement with the data
in the regions of the quasielastic peak, the dip region and the
$\Delta$ peak is good. On the other hand there is also agreement 
with the longitudinal and transverse response functions, particularly
with the latest analysis of \cite{JOU} taking the world data.

The study of \cite{AMPARO} was done in a covariant way and the response 
functions are associated to the virtual photon self-energy. Thus,

\be
   \begin{array}{l}
      W_{L}=\frac{\displaystyle{q^{2}}}{\displaystyle{\pi e^{2} |\vec{q}\,|^{2}}}
         {\displaystyle \int d^{3}r Im \Pi^{00}(q,\rho(\vec{r}\,))
            }\\
               \\
                  W_{T}=-\frac{\displaystyle{1}}{\displaystyle{\pi e^{2}}}{\displaystyle
                     \int d^{3}r Im \Pi^{xx}(q,\rho(\vec{r}\,))}
                        \end{array}
                           \ee

 Furthermore, the many body techniques used there, as it was the case
before in the study of the different reaction channels in pion-nucleus
\cite{LOR} and real photon-nucleus reaction \cite{CARRASCO}, allow one to
separate the different contributions to $Im \Pi^{\mu \nu}$
and associate them to the different reaction channels in the first step
of the reaction. In order to make the statement more explicit, let us
briefly discuss the origin and the separation of these contributions.
In ref. \cite{AMPARO} we introduced in a systematic way processes which
involve $1p1h$ excitation, $2p2h$ excitation, $3p3h$ excitation,
$1p1h1\pi$ excitation and $2p2h1\pi$ excitation. Some of the
Feynman diagrams which give rise to $Im \Pi^{\mu \nu}$ are drawn
in fig. 1.

The imaginary part from a Feynamn diagram is 
evaluated according to Cutkosky rules.  The rules \cite{ITZ}
essentially consist of the following: we cut the intermediate
states with a line at equal times and when all these states
are placed on shell in the integrations over the
internal variables, this contributes to $Im \Pi^{\mu \nu}$. Placing on shell 
a particle means that we take the imaginary part of its propagator.
For instance, for a pion

\be
 D(q) \rightarrow   2\,i\,\Theta(q^{0}) ImD(q)\,=\,2\,i\,\Theta(q^{0})\,(-\pi)\,
\delta(q^{2}-m^{2}_{\pi})
\ee

\noindent
and we see technically the meaning of placing the particle on shell.
This way of getting $Im \Pi^{\mu \nu}$ is doubly rewarding. First,
because of the economy versus the equivalent and more laborious way
(used in fact to prove the Cutkosky rules) of using Wick 
rotations or other alternative methods. Second, because placing some
intermediate particles on shell is telling us which channel of
the total cross section we are considering. Cutkosky rules 
are intimately associated to the optical theorem which relates
the total cross section to the imaginary part of the forward 
elastic scattering amplitude. Here it is the same: one is relating
the imaginary part of the self-energy (also forward) to the probability
of reaction. Then, the different cuts of intermediate states and 
their corresponding contribution to $Im \Pi^{\mu \nu}$, provide us 
with the physical channels of the reaction and the corresponding 
contribution of these channels to the reaction probability, 
respectively.

\vskip 0.2cm
\centerline{\protect\hbox{\psfig{file=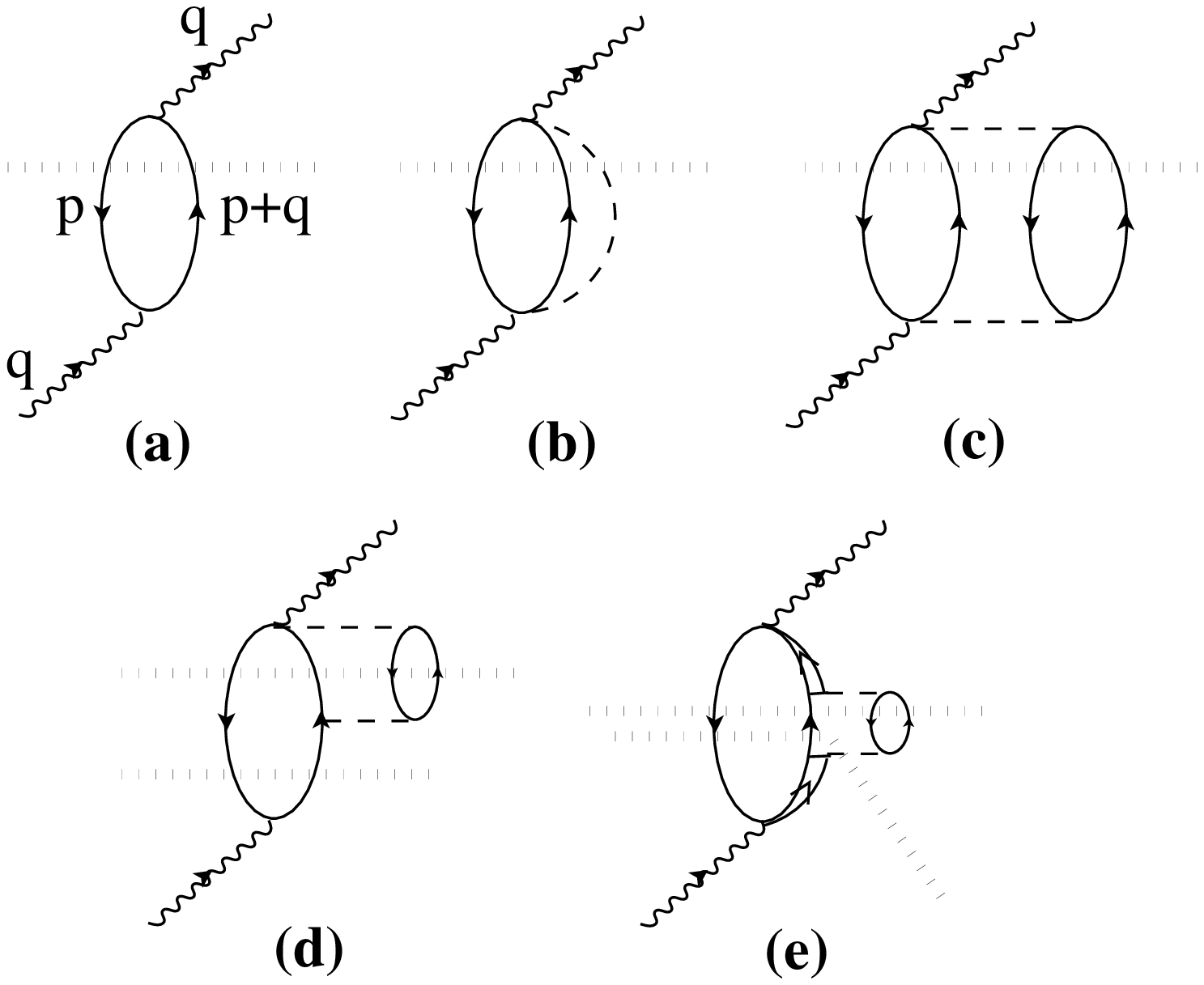,width=11.cm}}}
\vskip 0.2cm
\noindent
{\small {\bf Fig.1} Some many body Feynman diagrams whose imaginary parts contribute
to the total inclusive  $(e,e^{\prime}X)$ cross section. The dotted lines
indicate which particles are put on shell when evaluating the imaginary
part of each diagram.}
\vspace*{0.25cm}

If we look at fig. 1 we can see the different sources of imaginary part
represented by the cuts of the horizontal lines. Hence, in diagram (a)
the cut represents the channel of $1p1h$ excitation or absorption
of the photon by one nucleon. The reaction probability associated to
this diagram, or equivalently, its contribution to $W_L$ and $W_T$
is calculated in ref. \cite{AMPARO}. Here we would like on top of that to
have the distribution of momenta of the outgoing nucleon. This can be done
by not performing the integration over the momentum of the nucleon 
occupied states in \cite{AMPARO} since $p^{\prime}_{N}$ for the outgoing 
nucleon is $p+q$. However, since we generate events probabilistically
one by one, we choose the alternative procedure of generating a random
momentum from the local Fermi sea, and this will generate 
an outgoing momentum $p^{\prime}_{N}=p_{N}+q$
 ($p_N \equiv p$ in fig.1). If it happens that 
$|{\vec{p}\,}^{\prime}_{N}|<k_{F}(r)$ (the local Fermi momentum) then the event is 
Pauli blocked, it is dismissed and another event is generated. Thus, we have 
already the configuration of the final state after the first step: 
in this case just one nucleon produced in the point $\vec{r}$ of the nucleus
with momentum ${\vec{p}\,}^{\prime}_{N}$. 
As with respect to having a proton and a neutron in the final state, this is 
trivially done, since the contribution of diagram 1(a) in ref. 
\cite{AMPARO} was already splitted into a proton and a neutron
induced ones (see eq. (71) of ref. \cite{AMPARO}).

In diagram (b) the cut represents $1p1h1\pi$ excitation or, what is the same,
the first step in $(\gamma^{*},\pi)$. Here we have a probability for
this to occur, which is evaluated in \cite{AMPARO} and now one has 
to generate events with a momentum ${{p}}^{\prime}_{N}$ and
${{p}}_{\pi}$ for the outgoing $N$ and $\pi$. These
momenta are generated according to the weight that they have in the
evaluation of the probability, which actually involves integrations over 
these momenta. The technical way to do it is using the random procedure
which will be shown later on.

The diagram (c) has a cut corresponding to $2p2h$ excitation.
Here the channel represented by this cut is two nucleon emission in the
first step. Again we will have to determine the outgoing nucleon
momenta through a random procedure but weighed by the momenta 
distribution  implicit in the formulae which evaluate the probabilities 
by integration over these momenta.

In diagram (d) there is a novelty, peculiar to virtual photons. Indeed,
 the 
diagram has actually two sources of imaginary part, one of them corresponding
to the upper cut, where we have a $2p2h$ excitation and another one 
corresponding to the lower cut where we have $1p1h$ excitation. The lower cut
has the same structure as (a) and can be cast together using the structure
of (a) but with the upper vertex renormalized. In this case we have 
1 particle emission. However, the upper cut would contribute to $2N$ 
emission.

Finally, the diagram (e) contains two cuts, the upper one which corresponds
to $2p2h$ excitation, or $2N$ emission, and the lower one, which appears 
because we have a full (non- static) pion propagator, and which
corresponds to having $1p1h1\pi$ excitation, or equivalently, $1N$ and $1\pi$
emission.

These diagrams, and the corresponding distributions of pions and 
nucleons that we get from them, provide the configuration of the final particles
after the first step of the collision. Thus we have a set of nucleons and
pions with definite momenta produced at the point $\vec{r}$ of the nucleus. 
These particles still have to cross the nucleus before they are eventually
detected, but in their way out
the nucleus they can undergo different reactions with other nucleons, thus 
changing their momenta and leading to the ejection of more nucleons.
These secondary steps will be followed by means of a Monte Carlo simulation
method and will be discussed in forthcoming sections.  

\section{The Monte Carlo simulation}

\subsection{Generating and tracing the pions}

Let us assume that one pion is produced in a nuclear element of volume, $d^3r$.
Our simulation procedure allows us to decide (depending on the corresponding
 probabilities): first, via which process the pion has been produced
  and its position inside the nucleus. Second, the charge and momentum of the 
  produced pion.

  \vskip 0.2cm
      \centerline{\protect\hbox{\psfig{file=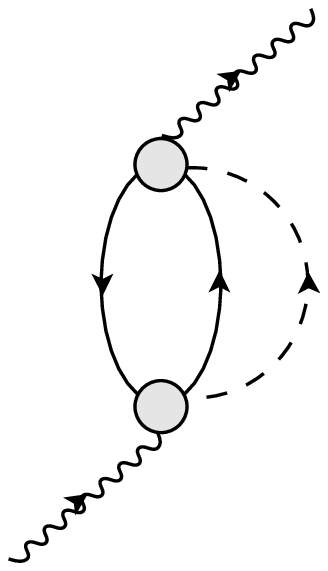,width=3.6cm}}}
      \vspace*{-4.7cm}
            \vskip 0.2cm
\noindent            
 {\small {\bf Fig.2} Lowest order contribution to the pion production 
 channel. The dot stands for the full $\gamma^{*} N \rightarrow N \pi$
 T-matrix, taken from the model of ref. \cite{AMPARO}.}
\vspace*{0.3cm}

  As we have shown in a previous paper \cite{AMPARO} the pion production channel has
  different sources: {\bf (a)} the contribution of the diagrams depicted
  in fig. 2, {\bf (b)} the quasielastic contribution of the $\Delta$ piece
  (fig. 3) and, finally, {\bf (c)} the contribution of the diagram depicted
  in fig. 4 which is related to the $(\gamma^{*},\pi \pi)$ channel.
  Obviously, as we discussed in \cite{AMPARO}, the main contribution
  corresponds to diagrams {\bf (a)}.

\vskip 0.2cm
\centerline{\protect\hbox{\psfig{file=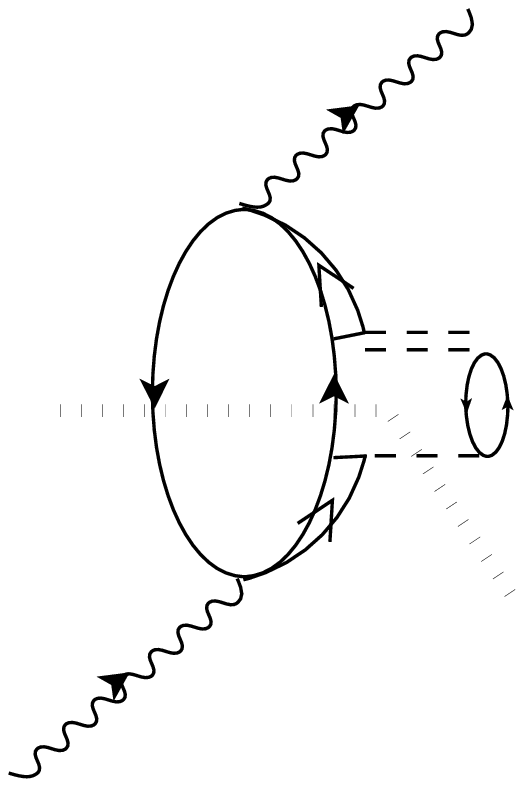,width=3.6cm}}}
\vskip 0.2cm
\noindent
{\small {\bf Fig.3} Contribution of the $\Delta$-self-energy in the nuclear medium
to the pion production channel. }
\vspace*{0.3cm} 
 
We evaluate the contribution of the diagrams {\bf (a)}
 by using the elementary cross section for
$\gamma^{*} N \rightarrow N \pi$ and taking into account nuclear 
medium effects: Fermi motion of the nucleons, modification of the
$\Delta$ width in the nuclear medium and the corresponding renormalization 
of the different terms.

\vskip 0.2cm
\centerline{\protect\hbox{\psfig{file=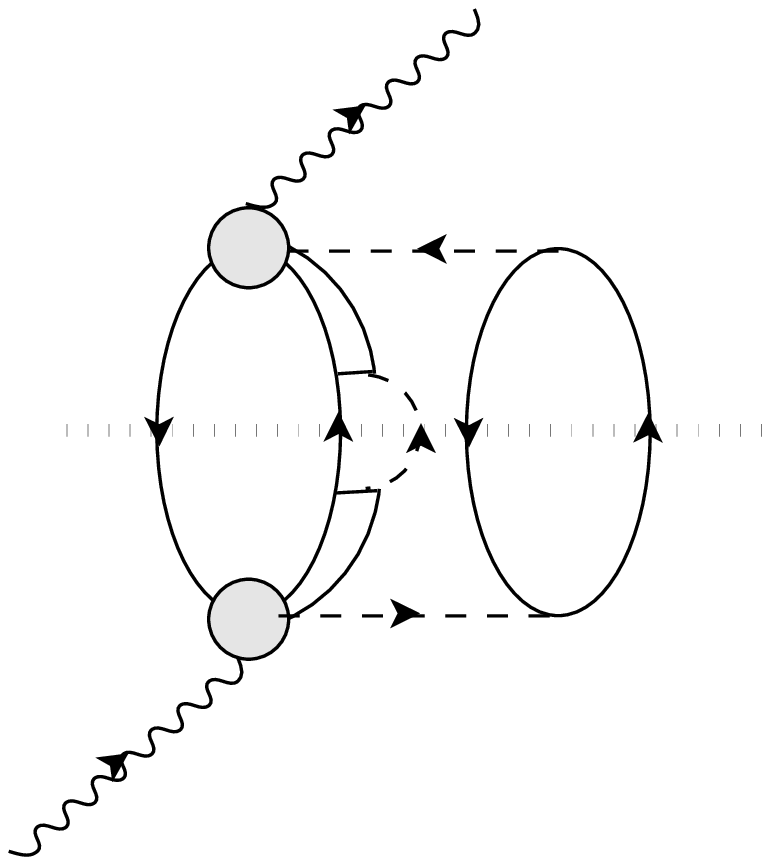,width=4.3cm}}}
\vskip 0.2cm
\noindent
{\small {\bf Fig.4} Contribution to the 2p2h1$\pi$ channel obtained from the
$\gamma^{*} N \rightarrow N \pi \pi$ T-matrix (full circle) used in ref.
\cite{AMPARO}.}
\vspace*{0.3cm}

In ref. \cite{AMPARO} we calculate $d\sigma/d\Omega^{\prime}dE^{\prime}$,
for the $(e,e^{\prime})$ reaction. We shall refer to it
as the $\gamma^*$ cross section. In particular the contribution of
the $(\gamma^* , \pi)$ channel from the diagrams {\bf (a)} 
will be given by:

\be
\begin{array}{ll}
\Frac{d{\sigma}^{\lambda}_{a} }{d\Omega^{\prime} dE^{\prime}}
= & {\displaystyle \int}d^{3}r\Frac{d^{3}p}{(2\pi)^{3}}
d\Omega^{*}_{CM}\Theta (cos\,\theta_{Max}-cos\,\theta_{CM} )
\Frac{|\vec{q}_{l}|}{|\vec{q}\,|}  
\times\\
 & \\
 & \times 2\left[n_{p}(\vec{p})
 \Frac{d\sigma (\gamma ^*p)^{\lambda}}{d\Omega^{\prime} dE^{\prime} d\Omega^{*}_{CM}}+
 n_{n}(\vec{p})\Frac{d\sigma (\gamma^* n)^{\lambda}}{d\Omega^{\prime}
 dE^{\prime} {d\Omega^{*}}_{CM}}\right]
 \end{array}
 \ee
      
\noindent
where $\int{d^{3}r}$ is extended to the nuclear volume; 
$\int{d^{3}p}$ corresponds to the integration over the nucleon momenta;
$n_p$ and $n_n$ are the proton and neutron occupation numbers;
$\Theta (cos\,\theta_{Max}-cos\,\theta_{CM} )$ is imposed by
Pauli blocking (see ref. \cite{LOREN}) and ${\Omega}^{*}_{CM}$ is the
solid angle of the pion in the CM of the $\gamma^* N$ frame.
 The index
$\lambda$ stand for the charge of the pion.
The factor $|\vec{q}_{l}|/|\vec{q}\,|$ is related to
the different flux of particles in the nuclear frame and in the nucleon frame
($|\vec{q}_{l}\,|$ is the photon lab momentum in the $\gamma^* N$ system
and $|\vec{q}\,|$, the momentum in the nuclear system)
and it is actually very close to unity \cite{LOREN}.
  
Contributions {\bf (b)} and {\bf (c)} mentioned before need at least
two nucleons and they present a $\rho^2$ dependence in the limit
of small densities. We calculate these contributions following the
procedure explained in ref. \cite{AMPARO}:

    \be
     \frac{\displaystyle{{d}^{2}\sigma^{\lambda}_{b,c}}}
      {\displaystyle{d\Omega^{\prime} dE^{\prime}}}=
        -\frac{\displaystyle{\alpha}}{\displaystyle{q^{4}}}\frac{
          \displaystyle{|\vec{k}^{\,\prime}|}}{\displaystyle{|\vec{k}|}}\frac{
            \displaystyle{1}}{\displaystyle{(2\pi)^{2}}}
              {\displaystyle\int d^{3}r
                \left(Im{\Pi}_{\gamma\,(b,c)}^{\mu\nu}L_{\mu\nu}\right)P_{\lambda}}
                \ee
                
\noindent
   where $P_{\lambda}$ is the probability related to the charge $\lambda$
 of the pion (i.e., the isospin dependence). For {\bf (c)}
 these probabilities are:
 
\be
   P_{0}({\pi}^{0})=\frac{\displaystyle{2}}
       {\displaystyle{12}}\,\,\,;\,\,\,P_{\pm}({\pi}^{\pm})
           =\frac{\displaystyle{5}}
                   {\displaystyle{12}}\,\mp\,\frac{\displaystyle{4}}
                      {\displaystyle{12}}\frac{\displaystyle{N\,-\,Z}}
                            {\displaystyle{A}}
                                \ee
                                
And for {\bf (b)} , we consider the probability corresponding to a resonant
reaction $(\gamma^{*}, \pi)$ with a renormalized pion (because this is the
main contribution):

\be
     P_{0}({\pi}^{0})=\frac{\displaystyle{2}}
 {\displaystyle{3}}\,\,;\,\,P_{+}({\pi}^{+})=
      \frac{\displaystyle{Z}}
        {\displaystyle{3A}}\,\,;\,\,
          P_{-}({\pi}^{-})=
       \frac{\displaystyle{N}}
        {\displaystyle{3A}}
       \ee               

   Once the pion is produced, one must decide which process took place 
 ({\bf (a)},{\bf (b)} or {\bf (c)}). If we write:
 
\be
 \frac{\displaystyle{ {d}^{2}{\sigma}^{\lambda} }  }                
   {\displaystyle{d\Omega^{\prime} dE^{\prime}}}=
                 \displaystyle{\sum_{j}\,
                      \frac{\displaystyle{ {d}^{2}{\sigma}^{\lambda}_j } }
                              {\displaystyle{d\Omega^{\prime}dE^{\prime}}}}
                              \ee

\noindent
where the sum over $j$ stands for the {\bf (a)}, {\bf (b)} and {\bf (c)} mechanisms
given in eqs. (3),(4). The choice of the process is made according
to these $d^2 \sigma^{\lambda}_{j}/d\Omega^{\prime} dE^{\prime}$.

The angular distributions $(\theta, \phi)$ for pions 
in cases {\bf (a)} and {\bf (b)} 
are generated according to the
differential cross section ${d\sigma (\gamma^{*} N \rightarrow \pi N)
}/{d\Omega^{\prime} dE^{\prime}d{\Omega}^{*}_{CM}}$ which appears in eq.(3),
while for the case of {\bf (c)}
 we consider
the momenta distributions given by three-body phase space.

As we can see from eqs.(3),(4), the spatial location of each event is 
easily generated, because the cross section is expressed in terms
of probabilities by unit of volume.

Let us show how the selection of $\vec{r}$ and angle is made
for the mechanisms {\bf (a)} and {\bf (b)} (in case {\bf (c)} would be
done analogously following the phase space distributions).
We define for each 
channel of charge:

\be
\begin{array}{rcl}
f(\vec{r}\,)\,\,\,&/&\,\,\,\Frac{d{\sigma}^{\lambda}_{(a),(b)}}
{d\Omega^{\prime} dE^{\prime}}=
\displaystyle{\int_{0}^{\infty} dr\,f(r)}\\
&&\\
g_r(cos\,\theta)\,\,\,&/&\,\,\, f(r)=\displaystyle{\int_{-1}^{+1}
 d(cos\theta)\,g_r(cos\,\theta)}\\
  &&\\
  t_{r,\theta}(\phi)\,\,\,&/&\,\,\, g_r(cos\,\theta)=
  \displaystyle{\int_{0}^{2\pi}
   d\phi\,t_{r,\theta}(\phi)}\\
    &&\\
    F(r)&=&\displaystyle{\int_{0}^{r}dr^{\prime}\,f(r^{\prime})}\\
    &&\\
    G(cos\,\theta)&=&\displaystyle{\int_{-1}^{cos\,\theta}
    d(cos{\theta}^{\prime})\,g_r(\cos{\theta}^{\prime})}\\
    &&\\
    T(\phi)&=&\displaystyle{\int_{0}^{\phi}
    d{\phi}^{\prime}\,t_{r,\theta}(\phi^{\prime})}
    \end{array}
    \ee
    
We generate three random numbers $x$,$y$,$z$ $\in\,[0,1[$ and choose 
one nuclear radius $r$ such that $F(r)=xF(\infty)$ and two
angles $\theta$ and $\phi$ such that $G(cos\,\theta)=y\,G(cos\,\theta=1)$
and $T(\phi)=z\,T(\phi=2\pi)$. In order to fix the pion kinematical
variables, a random momentum $\vec{p}_N$ is chosen from the Fermi sea.

Once the kinematical variables are fixed, they are used as input in the
computer simulation code for pion propagation described in refs.
\cite{RAFA,LOR} which gives us differential cross sections for every
pion channel.
The mentioned simulation takes into account the two possible kind of reactions 
in the way out of the pion: on the one hand, the pion can 
be absorbed, using its energy in 
ejecting two or more nucleons. On the
other hand, the pion can  ``suffer'' quasielastic collisions where the pion 
gives part of its energy to the nucleus.
In this latter case the pion can undergo charge exchange in one unit or remain
with the same charge. Through double quasielastic collisions 
one can also have pion double charge exchange.
 Details can be seen
in ref. \cite{LOR}. 
                                 
 \subsection{Nucleons: excitation mechanisms}
 
 Nucleons can be excited in different ways through induced reactions by
 a virtual photon:
 
   (i) {\underline {Direct $\gamma^{*}$-absorption}}:
   
   The virtual photon is absorbed and gives its energy to ${\cal{ N}}$ nucleons.
   For instance, the mechanism of fig. 1(a) leads to a primary
   one nucleon emission, while the mechanisms of fig. 1(c) leads to a primary 
   two nucleon emission.
   Details of this can be seen in ref. \cite{AMPARO}.
   
   (ii) {\underline {$(\gamma^{*},\pi)$ reactions}}:
   
   Above the electropion production threshold  
  the  $\gamma^{*} N\rightarrow \pi N^{\prime}$ reaction can occur. In this
   reaction (with the exception of a small fraction of coherent
    electropion production) one nucleon is excited. The excitation energy
    of the nucleon $N^{\prime}$  is small because most of the energy is used
    in creating the pion. Pauli blocking makes the events
    with more than one excited nucleon not too relevant.
    
     (iii) {\underline {$(\pi,\pi^{\prime})$ knock-out }}:
     
     Once a $\pi$ is produced, subsequent quasielastic collisions
     $N(\pi,\pi^{\prime})N^{\prime}$ are possible.
     
(iv)  {\underline { $\pi$ absorption or indirect $\gamma^{*}$-absorption}}: 
        
       The produced $\pi$ cannot only be multiply scattered but 
       also be absorbed by $ {\cal {N}}$  nucleons exchanging virtual pions or other mesons.
       In this case ${\cal {N}}$ must be greater than 1 and, in our approach, we
       consider events up to $ {\cal {N}}=3$.
       
  (v) {\underline{NN- collisions }}:
  
       The number of excited nucleons increases by means of their secondary
       collisions. A nucleon can share its energy with another one in such a way
       that both remain after the interaction with energy greater
       than the Fermi energy. These secondary nucleons will appear 
       mainly in the lower part of the spectrum. 
       
       Obviously, in order to explain the nucleon spectra, 
       one must consider all these processes simultaneously.
       
       Finally, in the next subsection we will discuss the mechanism 
       (v) above. 
       
       We discuss here briefly how the nucleons are generated in the first
       two mechanisms quoted above. Those  generated in the $(\pi, \pi^{\prime})$
       and $\pi$ absorption steps are discussed in ref. \cite{RAFA}. 
       
       \begin{itemize}
       
       \item{Direct $\gamma^{*}$-absorption}
       
       As we commented before, our theoretical approach allows us to determine
       which is the fraction of direct  $\gamma^{*}$-absorption corresponding
       to one, two or three nucleons. Furthermore, it provides us with the
       probabilities for the different channels in each point of the nucleus.
       The MC selects randomly -according to the distribution of probability-
       which process takes place and where does it happen. In section 2 we described 
       how the nucleon momenta and charge are generated for the 1$N$ induced
       $\gamma^*$ absorption (see fig. 1(a)). For the case of
       two and three nucleon induced $\gamma^*$ absorption 
        the momenta of the
       outgoing particles will be chosen according to phase space 
       distributions. Their charges will be selected according to the isospin
       dependence of the process. For example, the non-resonant terms (called
       background terms) in the absorption of the virtual photon 
       by two nucleons, correspond mainly to absorption by pairs 
       proton-neutron (pn)  because most of them
       imply the exchange of a charge pion between both nucleons.
       Hence they have a weight proportional to $\rho_n \rho_p$
       ($\rho_p$
       ($\rho_n$) is the nuclear density of protons (neutrons)). 
        In some of
       these reactions, a $\pi^0$ is exchanged and they lead to the
       absorption by pairs proton-proton (pp) and neutron-neutron (nn), thus having
       weigths ${\rho}^{2}_{p}$ and ${\rho}^{2}_{n}$ respectively
       as in the case of 
       the resonant term which has a contribution of two-body with an isospin
       dependence 
       
        $$
       \Frac{1}{6}{\rho}_{p}^{2}\,+\,\Frac{4}{6}{\rho}_{p}{\rho}_{n}\,+\,
                      \Frac{1}{6}{\rho}_{n}^{2}
                             $$ 

       \noindent 
       and a contribution of three-body which splits into the four possible
       channels according to the relatives probabilities

             $$
                   \Frac{1}{18}{\rho}_{p}^{3}\,+\,
                         \Frac{8}{18}{\rho}_{p}^{2}{\rho}_{n}\,+\,
                               \Frac{8}{18}{\rho}_{p}{\rho}_{n}^{2}\,+\,
                                     \Frac{1}{18}{\rho}_{n}^{3}
                                           $$

       \item{Direct ($\gamma^{*},\pi$)}
       
       Once the virtual photon momentum is fixed,
       our code selects a random nucleon from the Fermi sea and chooses
       the charge and the momentum of the outgoing pion according to the 
       nuclear cross section (see section 3.1). With this, all the properties of the excited nucleon
       are also fixed.
       
       \end{itemize}

\subsection{Nucleon propagation}

       The nucleons in the nucleus move under the influence 
       of a complex optical potential. The imaginary part
       of the potential is related to the probability of nucleon quasielastic
       collisions in the nucleus (and extra pion production at higher energies
       which we do not consider here). We consider explicitly these collisions
       since they generate new nucleons going outside the nucleus. As with respect
       to the real part, we use it to determine the classical trajectories
       that the nucleons follow in the nucleus between collisions.
       
       As done in ref. \cite{RAFA} we take as the real part of the nucleon-nucleus
       potential

  $$
     V(r)=V_{\infty}-{\cal{E}} (r)=-\Frac{k_{F}^{2}}{2M_{N}}=
     -\Frac{1}{2M_N}
        \left(
           \frac{3}{2}\pi^{2}\rho (r)\right)^{2/3}
              $$

It represents the interaction of a single nucleon with the average 
potential due to the rest of the nucleons. This choice of $V(r)$ means that 
the total nucleon
energy is the difference between its kinetic energy and the Fermi
energy, $\epsilon_F$.

As with respect to collisions 
in our Monte Carlo simulation we follow each excited nucleon by
 letting it move
a short distance $d$ such that $Pd<<1$ ($P$ represents the probability
per unit of length for a quasielastic collision ). The new position 
(${\vec{r}\,}^{\prime}$)                     
  and momentum (${\vec{p}\,}^{\prime}$) are taken from the Hamiltonian
  equations as
  
  $$
  \left\{
      \begin{array}{l}
            {\vec{r}\,}^{\prime}=\vec{r}+\delta\vec{r}=\vec{r}+\hat{p}d \\
           {\vec{p}\,}^{\prime}=\vec{p}+\delta\vec{p} \,\,;\,\, 
           \delta\vec {p}\,= - \Frac{\partial V}{\partial r} 
           \Frac{M_N d}{p} \hat{r}
                         \end{array}
                          \right.
                           $$
                           
\noindent
which follow from the Hamilton equations or equivalently from  
                      energy and angular momentum conservation. 
                        
Our code selects randomly, according to the reaction probabilities 
which will be discussed in 3.4, if the nucleon is 
scattered or not and, in the case
of scattering, what kind of process takes place.
If no collision takes place, we move the nucleon again.
When the nucleon leaves the nucleus we stop the process and it is counted
as a contribution to the cross section. If a NN scattering is selected
instead, 
we take
a random nucleon from the Fermi sea and calculate the initial kinematical
variables ($P^{\mu}$ and s, full four-momentum of the nucleon-nucleon system
in the nuclear frame and invariant energy, respectively). Then, a 
$cos\,\theta_{CM}^{N^{\prime}}$ is selected, according to the
 expression given in ref. \cite{RAFA}. 
 This expression gives us
 the correct probability given by $d\sigma^{NN^{\prime}}/d\Omega_{CM}$   
 (in the same appendix) plus Pauli blocking restrictions. We take
 also into account Fermi motion and renormalization effects in the angular 
 dependence. We take them into account by multiplying each event by a weight
 factor

$$
\xi =\lambda (N_{1},N_{2})\hat{\sigma}^{N_{1}N_{2}}\rho_{N_{2}}
$$

\noindent
where $1/\lambda(N_{1},N_{2})$ is the probability by unit of length that  
one nucleon $N_1$ collides with another nucleon
$N_2$. In average, this factor $\xi$ is equal to one. 

Our method assumes that the nucleons propagate semiclasically in the nucleus.
The justification of this hypothesis for real photons is given in ref.
\cite{RAFA}. In the next section we give some detail on the evaluation
of the equivalent NN cross section in the medium.

       \subsection{NN cross sections}
       
          We are using the parameterization of the NN elastic cross 
          section given in the appendix of ref. \cite{RAFA}. Due to the
          fact that for particles of low momenta, the Monte 
          Carlo induces large errors, we are not considering collisions
          of nucleons with kinetic energies below 50 $MeV$.
          
On the other hand, the reaction probability will change due to the nuclear medium
effects (Fermi motion, Pauli blocking and medium renormalization). Then,
according to ref. \cite{RAFA}, the expression for the
mean free path ($\lambda$) of the nucleon is given by

\be
\Frac{1}{\lambda}(N_1)
= 4 {\displaystyle \int}\Frac{d^{3}p_{2}}{(2\pi)^{3}}n_{p_2}
\left[
\Frac{Z}{A}\hat{\sigma}^{N_{1}p}(s)+
\Frac{(A\,-\,Z)}{A}\hat{\sigma}^{N_{1}n}(s)\right]
\Frac{|\vec{p}_{1\,lab}|}{|\vec{p}_{1}|}
\ee

\noindent
with $P$, the full four-momentum of the NN system in the nuclear frame;
$s=P^{2}=(p_1+p_2)^{2}$; $N_1$, the outgoing nucleon and  $N_2$, the nucleon
in the medium.                      
       
Furthermore,

\be
\hat{\sigma}^{N_{1}N_{2}}
= {\displaystyle \int}d\Omega_{CM}\Frac{d\sigma^{N_{1}N_{2}}}
{d\Omega_{CM}}f(q)\Theta(\kappa-|\hat{P}.\hat{p}_{CM}|)
\ee

\noindent 
where CM is the NN center of mass frame and

 \be
  \kappa
   = x\Theta(1-|x|)\,+\,\Frac{x}{|x|}\Theta(|x|-1)
    \ee
    
On the other hand

 \be
   x
      = \Frac{P^{0}\,p^{0}_{CM}-\epsilon_{F}\sqrt{s}}
         {|\vec{P}\,||\vec{p}_{CM}|}
             \ee
             
\noindent
where $\vec{p}_{CM}$ is the nucleon momentum in the CM frame and $\epsilon_{F}$
is the Fermi energy. The function $f(q)$ corresponds to

\be
f(q)
= \Frac{1}{\left|1-U(q)\left(\frac{{f}_{\pi NN}}{m_{\pi}}\right)^{2}
V_{t}(q)\right|^{2}}
\ee

\noindent
where $U(q)$ is the Lindhard function ($U(q)=U_N+U_{\Delta}$) and $q$ is the
momentum transfer in the nuclear frame.
$V_t$ is the spin-isospin effective interaction (details of this can be seen
in ref. \cite{AMPARO}). In these expressions,
$\Theta(\kappa-|\hat{P}.\hat{p}_{CM}|)$ takes into account Pauli blocking 
and $f(q)$, the medium renormalization.

\vspace*{1.8cm}

\section{Results}

\subsection{$(e,e^{\prime}N)$ reactions.}

   We show our results for the cross section of the $(e,e^{\prime}p)$ process
   as a function of the missing energy (which is defined as $E_m=\omega-T_p$,
   the difference between the energy transfer and the proton kinetic energy.
   
   In figs. 5-9 we show results for different kinematical configurations.
   The agreement with experiment is fair globally, although some punctual
   discrepancies appear for certain pieces of data. The trend of the
   spectrum is well reproduced, with the cross section increasing with the 
   missing energy in the region of energies studied. The region of large
   missing energy comes from pion production, with or without reabsorption,
   and from secondary nucleons originated in the $NN$ collisions. Some
   of these mechanisms, however, are more visible in the two nucleon 
   emission as we show later.
   
   Further results can be seen in ref.\cite{Raul} which show a very good
    agreement 
   between experiment and the present theory. 
   
       In fig. 10 we show a missing energy spectrum for pn emission at
       $\omega=400$ MeV. We see two peaks, one at low energies and the
       other at energies around 250 MeV. In the figure we have also separated
       the different sources of contribution. The peak at low energies
       corresponds to the original $\gamma^*$ absorption by a pair 
       of nucleons. We include there all events corresponding to the case 
       where the pairs leave the nucleus without further interaction as well
       as those events where any of the nucleons undergoes secondary 
       collisions.
             We also include the contribution from original three
               nucleon $\gamma^*$ absorption, which in this case is
         relatively small. 
         
            The broad peak at high missing energies in fig.10 corresponds to
            original pion production. In this case one has the possibility
            that the pion is reabsorbed, which accounts for the largest part
            of the strength, while there is a fraction, wich increases 
            with the missing energy, corresponding to the case where the
            pion escapes from the nucleus.
            
            In fig. 11 we show the same figure for pp emission. The features
            are similar, only in this latter case the peak at low energies 
            corresponding to primary $\gamma^*$ absorption by a pair of
            nucleons (mostly pp) with or without FSI is about 5 to 6 times
            smaller than in the pn emission case. On the other hand, the
            broad peak at large missing energies corresponding to primary
            pion production is about 1/3 the size of the corresponding 
            pn peak, much as it happened in real $\gamma$ absorption.

\vspace*{1.9cm} 

   \centerline{\protect\hbox{\psfig{file=carbon0junio.ps,width=13.5cm}}}
    \vskip 0.2cm
\noindent    
   {\small {\bf Fig.5} The $^{12}C(e,e^{\prime}p)$ differential cross
   section as a function of the missing energy. 
  $E_e= 460$ MeV, 
   $\omega= 275$ MeV, $|\vec{q}\,| =401$ MeV  and
   $\theta_p=0^0$. 
    Experimental data from \cite{BAG}.}

\newpage
   \vspace*{0.4cm}
      \centerline{\protect\hbox{\psfig{file=carbon38junio.ps,width=12.5cm}}}
          \vskip 0.2cm
\noindent          
    {\small {\bf Fig.6}
    Same as fig.5 with   $E_e= 478$ MeV, 
    $\omega= 263$ MeV, $|\vec{q}\,| =303$ MeV and
          $\theta_p=38^0$.
              Experimental data from \cite{JAN2}.
    }

                \vspace*{0.7cm}
        \centerline{\protect\hbox{\psfig{file=carbon113junio.ps,width=12.5cm}}}
            \vskip 0.2cm
\noindent            
   {\small {\bf Fig.7}
   Same as fig.5 with 
$E_e= 478$ MeV,   
 $\omega= 263$ MeV, $|\vec{q}\,| =303$ MeV and
         $\theta_p=113^0$.
             Experimental data from \cite{JAN2}.
   }

\newpage

   \vspace*{0.4cm}
      \centerline{\protect\hbox{\psfig{file=carbonmenos80junio.ps,width=12.5cm}}}
          \vskip 0.2cm
\noindent          
  {\small  {\bf Fig.8}
   Same as fig.5 with $E_e= 509$ MeV, $\omega= 310$ MeV,
$|\vec{q}\,| =343$ MeV   
 and
            $\theta_p=-80^0$.
                         Experimental data from \cite{ZON}.
 }

   \vspace*{0.7cm}
         \centerline{\protect\hbox{\psfig{file=carbonmenos120junio.ps,width=12.5cm}}}
                   \vskip 0.2cm
\noindent 
    {\small {\bf Fig.9}
      Same as fig.5 with 
$E_e= 509$ MeV, $\omega= 310$ MeV,
$|\vec{q}\,| =343$ MeV       and
                  $\theta_p=-120^0$.
                   Experimental data from \cite{ZON}.
     }

\newpage

   \vspace*{0.7cm}
          \centerline{\protect\hbox{\psfig{file=pnjunio.ps,width=12.cm}}}
                 \vskip 0.2cm
\noindent                 
                  {\small {\bf Fig.10}
                      $^{12}C(e,e^{\prime}pn)$ differential cross section as 
         a function of the missing energy $E_m=\omega-T_{p1}-T_{p2}$. 
    $E_e= 705$ MeV, $\omega= 400$ MeV and
    $\theta_e =33.6^0$.     
          We have separated the different contributions: 2N+FSI
          (main histogram at low miss. energies), 3N+FSI 
          and
          1$\pi$+FSI (main histogram at high miss. energies).}

   \vspace*{0.2cm}
          \centerline{\protect\hbox{\psfig{file=ppjunio.ps,width=12.cm}}}
                 \vskip 0.2cm
\noindent                 
                  {\small {\bf Fig.11}
                      $^{12}C(e,e^{\prime}pp)$ differential cross section 
                 as a function of the missing energy.
                       $E_e= 705$ MeV, $\omega= 400$ MeV and
          $\theta_e =33.6^0$.            
                   We have separated the different contributions: 2N+FSI
                    (main histogram at low miss. energies),
                  3N+FSI
                             and
       1$\pi$+FSI (main histogram at high miss. energies).
                 }

  \newpage
  
In fig. 12 we show results for the $(e,e^{\prime}\pi)$ differential cross
section as a function of the pion momentum for two different angles. This
corresponds to pions which manage to scape the nucleus without absorption,
although they are allowed to undergo quasielastic collisions. The pion
momenta peak around the resonance region where pion production has 
its maximun strengh.

       In fig. 13 we show a picture in which we separate the primary
       pions produced into those which are absorbed and those which
       manage to escape as a function of $r$, the radius where they are
       produced. As can be seen, the pions which are produced in the interior
       of the nucleus are those which have more chances to be absorbed.
       The results are similar for other nuclei, except that the fraction
       of absorbed pions increases with A (see ref. \cite{tesis}).

   \vspace*{0.7cm}
         \centerline{\protect\hbox{\psfig{file=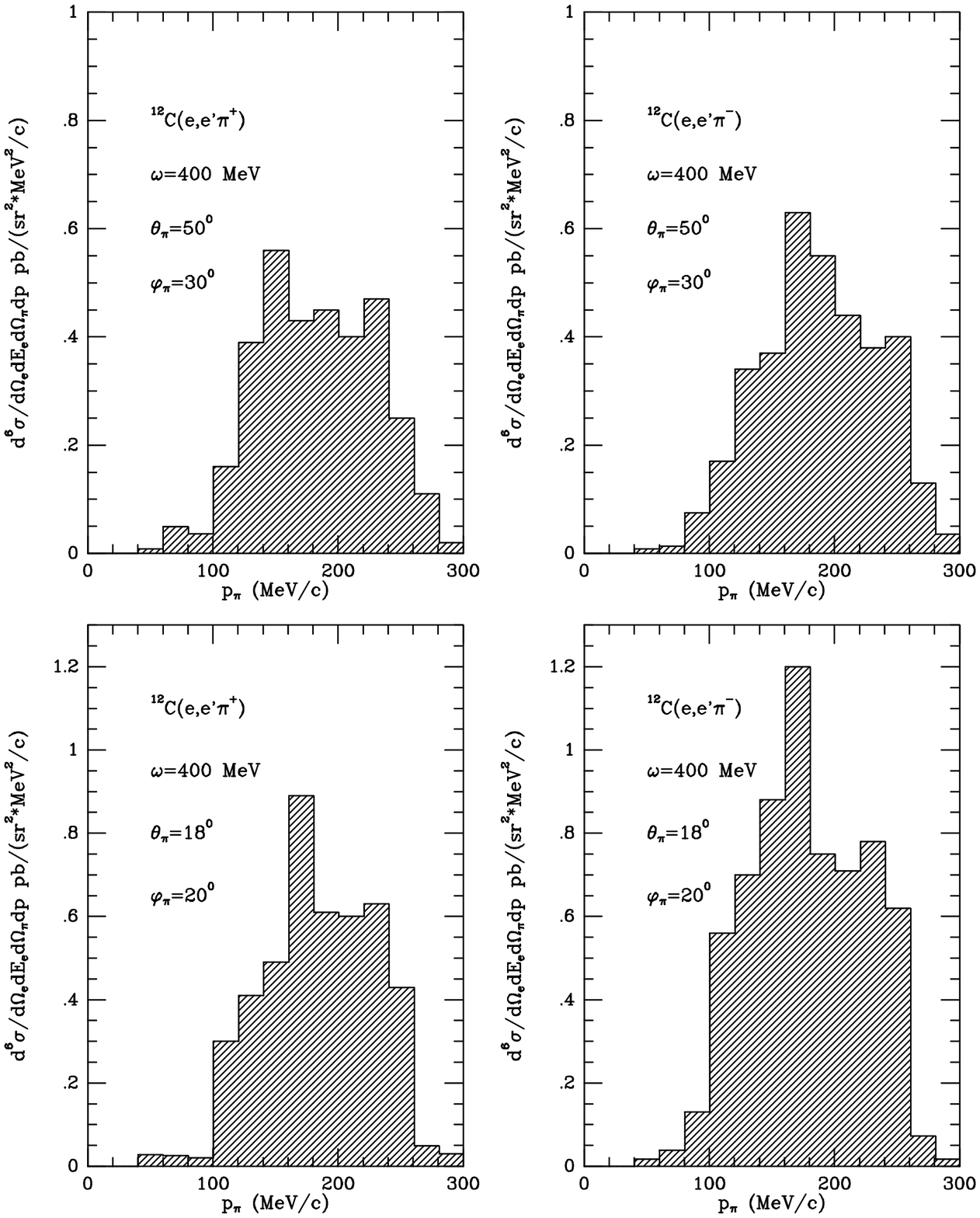,width=12.cm}}}
                   \vskip 0.2cm
\noindent
     {\small {\bf Fig.12}
     $^{12}C(e,e^{\prime}\pi)$ differential cross section for 
     two different angles and
      $E_e= 705$ MeV, $\omega= 400$ MeV and
          $\theta_e =33.6^0$.
                  }
                                
\newpage
         \centerline{\protect\hbox{\psfig{file=radiocarbon.ps,width=12.cm}}}
                            \vskip 0.2cm
\noindent                                 
                                 {\small {\bf Fig.13} 
           $^{12}C$: separation 
of primary pions into those which are absorbed (lower histogram) 
and those which manage to escape (upper histogram) as a function of r. 
 $E_e= 705$ MeV and  $\theta_e =33.6^0$.
  }

\vspace*{0.15cm}
  In figs. 14,15 we show the same results but now integrated over r, as 
  a function of $\omega$.
  Figure 14 shows that in  $^{12}C$ about one fourth of the primary pions
  produced are absorbed in their way out of the nucleus and the rest manages
  to escape. The situation is reversed in $^{208}Pb$, as can be seen in fig. 15,
  where about two thirds of the primary pions produced are absorbed
  in the nucleus and onlu one third escapes.

     \centerline{\protect\hbox{\psfig{file=carbonpions.ps,width=12.cm}}}

\noindent                                
   {\small {\bf Fig.14} Same as fig. 13 integrated over r and as a function
   of $\omega$.  $E_e= 705$ MeV and $\theta_e =33.6^0$.

\newpage                                                     

     \centerline{\protect\hbox{\psfig{file=plomopions.ps,width=12.cm}}}
                                      \vskip 0.2cm
\noindent                                      
     {\small {\bf Fig.15} Same as fig. 14 for $^{208}$Pb.
     $E_e= 705$ MeV and $\theta_e =33.6^0$.

\vspace*{0.3cm}                                                                                                  
In fig. 16 we show the cross section for the emission of a proton and
a pion in coincidence, as a function of $T_p$. We observe that 
the energies of the nucleons are relatively low, since 
the pion carries in this case  most of the energy. At higher energies,
 the proton energy distribution becomes obviously more spread.

   \vspace*{0.7cm}
       \centerline{\protect\hbox{\psfig{file=protonpluspion.ps,width=12.cm}}}
       \vskip 0.2cm
\noindent       
 {\small {\bf Fig.16}
    $^{12}C(e,e^{\prime}p)$ differential cross section plus
          a pion detected.
$E_e= 478$ MeV,
    $\omega= 263$ MeV, $|\vec{q}\,| =303$ MeV  and $\theta_p= 111^0$.        
  }
                                                                                     
\section{Conclusions}

We have carried out a systematic study of different exclusive channels in the 
$(e,e^{\prime})$ reaction, like $(e,e^{\prime} N)$, $(e,e^{\prime}NN)$,
$(e,e^{\prime} \pi)$ and $(e,e^{\prime} N \pi)$.
The procedure used to deal with this multichannel problem in the final state
of the $(e,e^{\prime})$ reaction, has been a Monte Carlo simulation method
in which the probabilities for the different steps have been evaluated
microscopically before using quantum mechanical many body techniques.

The procedure allows one to trace back to the primary step and 
subsequent secondary steps the different contributions 
to a certain final state channel. For instance, in pn emission 
the pn final state can come 
from direct knockout of a pn pair by the virtual photon. The same process
with final state interaction of some of the nucleons, from three body
knockout, with or without final state interaction of the nucleons. 
Also, from pion production in the primary step followed by pion reabsorption
of the pion by pairs or trios of particles.
This last process shows a strong A dependence in which the fraction of pions
which are absorbed changes from 1/4 to 2/3 from  $^{12}C$ to 
 $^{208}Pb$. This implies a pion absorption fraction proportional 
 to  $A^{\alpha}$, with $\alpha \approx 1.3$, while in pion absorption
 experiments, with cross sections close to $\pi R^2$, the absorption cross
 section goes as $A^{\alpha}$ with $\alpha=0.66$. This means that the indirect
 way of virtual photon absorption discussed above, with pion production plus
 reabsorption, is more sensitive to the intrinsic probability of pion
 absorption than the pion absorption cross section in pion nucleus
 collisions. In other words, one can learn more about the mechanisms
 of pion absorption studying indirect photon absorption (with
 real or virtual photons) than from pion absorption in pion nucleus 
 experiments.
 
 We have compared with a limited amount of experimental data here and
 found good agreement with experiment. Some experimental groups
 are comparing their data with the predictions
 of the present theoretical approach with good agreement so far.
 
 The value of the present work can be seen form different points of view.
  From one side, the comparison with data in different kinematical regions and different channels 
  can serve to assess the strengh of the different mechanisms discussed
  here. Conversely, one can use the present theoretical tool as a prospective device in order
  to find the kinematical situations which make it more efficient the 
  study of certain mechanisms that one choose to investigate.
  A similar tool for real photons \cite{RAFA} has been used efficiently 
  in both directions \cite{GROSS,GRAB,HELH} and it looks clear that
  the present study for virtual photons can play a similar role in the future
  in order to extract the maximum information available form forthcoming
  experiments at intermediate energy labs.
  
  \vspace*{0.8cm}
  Acknowledgments
  
       We would like to acknowledge useful discussions with R. Edelhoff
       and G. Rosner, J. Ryckebusch and J. Segura. 
       One of us, J. Nieves, acknowlodges to DEGS under contract PB95-1204.
       This work has been
       partially financed by CICYT contract number, AEN 96-1719.

\end{document}